\documentclass[useAMS,usenatbib]{mn2e}
\usepackage{graphicx,amssym,hyperref}
\def\oii{[{O~\sc ii}]}

\def\mgii{{Mg~\sc ii}}

\def\ha{H$\alpha$}
\def\hda{H$\delta_{\rm A}$}

\def\Wha{W(H$\alpha$)}
\def\d4n{D$_n$4000} 

\newcommand{\bc}{\begin{center}}
\newcommand{\ec}{\end{center}}
\newcommand{\cf}{\ifmmode C_f\else $C_f$\fi}

\title[Post-Starburst Galaxies in SDSS-IV MaNGA]
      {Post-Starburst Galaxies in SDSS-IV MaNGA}

\author[Chen et al.]{
\parbox[t]{\textwidth}{\raggedright
Yan-Mei Chen$^{1,2}$\thanks{Email: chenym@nju.edu.cn},
Yong Shi$^1$,
Vivienne Wild$^2$,
Christy Tremonti$^3$,
Kate Rowlands$^4$,
Dmitry Bizyaev$^{5,6}$,
Renbin Yan$^7$,
Lihwai Lin$^{8}$,
Rog\'erio Riffel$^{9,10}$
}\\
\vspace*{6pt}\\
$^1$Department of Astronomy, Nanjing University, Nanjing 210093, China\\ 
    Key Laboratory of Modern Astronomy and  Astrophysics (Nanjing University), Ministry of Education, Nanjing 210093, China\\
    Collaborative Innovation Center of Modern Astronomy and Space Exploration, Nanjing 210093, China\\
$^2$School of Physics and Astronomy, University of St Andrews, St Andrews, KY16 9SS, U.K.\\
$^3$Department of Astronomy, University of Wisconsin-Madison, 1150 University Ave, Madison, WI 53706, USA\\
$^4$Department of Physics \& Astronomy, Johns Hopkins University, Bloomberg centre, 3400 N. Charles St., Baltimore, MD 21218, USA\\
$^5$Apache Point Observatory and New Mexico  State University, P.O.  Box 59, Sunspot, NM,  88349-0059,  USA\\
$^6$Sternberg Astronomical Institute, Moscow State University, Moscow, Russia\\
$^7$Department of Physics and Astronomy, University of Kentucky, 505 Rose Street, Lexington, KY 40506-0055, USA\\
$^8$Institute of Astronomy and Astrophysics, Academia Sinica, Taipei 106, Taiwan\\
$^9$Departamento de Astronomia, Instituto de F\'isica, Universidade Federal do Rio Grande do Sul, Campus do Vale, 91501-970 Porto \\
Alegre, Brasil \\
$^{10}$Laborat\'orio Interinstitucional de e-Astronomia, 77 Rua General Jos\'e Cristino, 20921-400, Rio de Janeiro, Brasil\\}        
\begin{document}



\maketitle

\label{firstpage}

\begin{abstract}
  Post-starburst galaxies, identified by their unusually strong Balmer absorption lines and weaker than average emission lines, have traditionally been selected based on their central stellar populations. Here we identify 360 galaxies with post-starburst regions from the MaNGA integral field survey and classify these galaxies into three types: 31 galaxies with central post-starburst regions (CPSB), 37 galaxies with off-center ring-like post-starburst regions (RPSB) and 292 galaxies with irregular post-starburst regions (IPSB). Focussing on the CPSB and RPSB samples, and comparing their radial gradients in \d4n, \hda\ and \Wha\ to control samples, we find that while the CPSBs have suppressed star formation throughout their bulge and disk, and clear evidence of rapid decline of star formation in the central regions, the RPSBs only show clear evidence of recently rapidly suppressed star formation in their outer regions and an ongoing central starburst. The radial profiles in mass-weighted age and stellar $v/\sigma$ indicate that CPSBs and RPSBs are not simply different evolutionary stages of the same event, rather that CPSB galaxies are caused by a significant disruptive event, while RPSB galaxies are caused by disruption of gas fuelling to the outer regions. Compared to the control samples, both CPSB and RPSB galaxies show a higher fraction of interactions/mergers, misaligned gas or bars that might be the cause of the gas inflows and subsequent quenching. 
\end{abstract}


\section{Introduction}
\label{sec:intro}

The galaxy population in the local Universe is characterized by a clear bimodality in the 
color magnitude diagram \citep{baldry04, jin14}, with a ``blue cloud'' (the gas-rich star forming galaxies) and ``red 
sequence'' (gas poor quiescent galaxies). In between lies a minor population 
of ``green valley'' galaxies. This bimodality has built up over cosmic time, with galaxies migrating from the blue cloud to the red sequence \citep{bell04, brown07}. The low galaxy number density in the green valley has been argued to imply that the  
migration is relatively rapid \citep{martin07}, however, the exact timescales are still under debate \citep[for a recent summary see][]{rowlands18}. 

Post starburst (PSBs) galaxies, also 
called E+A/K+A galaxies, show unusually prominent Balmer absorption lines indicating an excess 
contribution to their light from intermediate-age stars (A- or F-type stars). This can be due to a recent 
starburst or rapid decline in star formation. Many such systems also show weak or absent nebular emission lines, 
implying an absence of the hotter, younger stars (O and B type). This implies an abrupt termination of the star 
formation (SF) process \citep{dressler83, poggianti99, goto03b}. Many have postulated that PSBs are observed 
in a short-lived transition phase, providing an evolutionary link between the blue cloud and red sequence 
galaxies \citep[e.g.,][]{yang04, yang06, kaviraj07, wild09, yesuf14, cales15, alatalo16a, wild16, rowlands18}, but whether this is true for 
most of PSBs remains unclear.  In this paper we will use the term ``quenching” to refer to a recent rapid shut-off of star formation 
and “quenched” to refer to a galaxy or region of a galaxy with minimal current star formation, regardless of whether this is a temporary 
or permanent state.  A quenched galaxy by this definition, may subsequently form stars.

Since their discovery PSB galaxies have been extensively studied. Significant progress was made with the advent of high quality multi object spectroscopy, in particular with the SDSS \citep{york00} surveys, which identified PSB galaxies based on light from the central 3'' diameter \citep{goto03a, goto05, goto07b, yan09}. The morphology of PSB galaxies is found to be
typically bulge dominated and sometimes with an underlying disk \citep{tran04, quintero04, goto05, wong12, maltby18, pawlik18}.  Disturbed morphologies or tidal features are present in many cases \citep[e.g.,][]{zabludoff96, goto05, yang08, 
yamauchi08, lin10, sell14, pawlik18}, although fade rapidly with starburst age \citep{pawlik16,pawlik19}. The disturbed, bulge-dominated morphologies, as well as the elemental abundances \citep{goto07a}, are consistent with the hypothesis that many of them are remnants of mergers or interactions, and the progenitors of early-type galaxies. 

Much attention also has been given to PSB galaxies in dense environments, where the environment could give rise to their sudden change in star formation activity. Studies of large galaxy clusters indicate an enhanced fraction of PSB galaxies compared to the field, at least in the cluster cores  \citep{dressler99,poggianti99,tran03, tran04, vonderlinden10, socolovsky18, paccagnella19}, which implies environmental processes such as ram-pressure stripping or harassment may be at play. Mergers are not expected to be important in clusters due to the high relative velocities of the galaxies. However, it is important to realise that in the local Universe the vast majority of PSB galaxies are found in the field, with little trend with local density \citep{quintero04,balogh05,goto05,hogg06,yan09,pawlik18}. This is again consistent with the majority of PSBs originating from mergers of star-forming galaxies, although additional processes may play a role in clusters. 

While many PSB galaxies possess both morphological and spectroscopic signatures of an evolutionary transition, the significant reservoirs of cold gas found in recent studies \citep{rowlands15,french15,alatalo16b} have raised some doubt as to this simple interpretation. Additionally, the EAGLE cosmological hydro-dynamic simulation reveals multiple evolutionary pathways for PSB galaxies in the local (simulated) Universe \citep{pawlik19}. Further studies are clearly needed to understand the origin and fate of these intriguing galaxies. 

In particular, the stellar population gradients and internal kinematics are poorly understood. In the last few years small numbers of PSB galaxies have been studied with long-slits or integral-field units (IFU) \citep{swinbank11, pracy10, pracy13, pracy14,hiner15}. Such studies are crucial to answer the most important question of active research: Do the stellar population gradients and kinematics of PSB galaxies support the picture of galaxy mergers forming spheroids, or are they consistent with models in which star formation is abruptly quenched in otherwise normal disk galaxies? The small samples, limited spectral or spatial coverage, and poorer data quality of these studies compared to the SDSS surveys, have thus far precluded a comprehensive answer to this question. 

Thanks to large sample of 4633 galaxies observed in the first three years of the fibre-optic IFU survey ``Mapping Nearby Galaxies at Apache Point Observatory'' (MaNGA), 
for the first time, we are able to search for PSB regions within the complete galaxy area. We can identify off-centre PSB regions,  ask what the difference is between these galaxies and galaxies with central PSB regions, and investigate the implications for their formation. In this paper we focus on galaxies with central and ring-like PSB regions. In Section 2, we introduce the MaNGA survey, sample selection and data analysis methods. The properties of the galaxies with PSB regions, including the host galaxy morphology, stellar populations and internal kinematics, are studied in Section 3. We discuss the observational results in Section 4 and a short summary is presented in Section 5. We use the cosmological parameters $H_0=70~{\rm km~s^{-1}~Mpc^{-1}}$,
$\Omega_{\rm M}=0.3$ and $\Omega_{\Lambda}=0.7$ throughout this paper.

\section{DATA}
\subsection{The MaNGA Survey}
MaNGA is one of three major programs of the ongoing fourth-generation Sloan Digital Sky 
Survey \citep[SDSS-IV;][]{bundy15, drory15, law15, law16, yan16a, blanton17, yan16b}, using 2.5m Sloan 
Foundation Telescope \citep{gunn06} at Apache Point Telescope (APO). MaNGA employs 
dithered observations with 17 fiber-bundle IFUs with 5 sizes vary between 19 and 127 
(or $12.5\sim32.5^{\prime\prime}$ diameter in the sky), depending on the apparent size of the 
target. Two dual-channel BOSS spectrographs \citep{smee13} provide simultaneous spectral 
coverage over 3622$-$10354\AA\ at $R\sim$2000. Between 2014 and 2020, MaNGA will 
have obtained IFU observations of $\sim$10,000 galaxies ($z \le$ 0.1) with stellar mass 
$M_*\ge10^9M_\odot$ and with an approximately flat $M_*$ distribution \citep{wake17}. 
MaNGA will observe 2/3 of the galaxy sample out to $\sim$1.5$R_{\rm e}$ and the other 1/3 to 
$\sim$2.5$R_{\rm e}$. As described by \citet{yan16a}, with a typical integration time of 3 hours, 
MaNGA reaches a signal-to-noise ratio (S/N) of $4\sim8$ per fiber in the $r-$band at a surface 
brightness of 23 ABmag arcsec$^{-2}$, which is the typical case for the outskirts of MaNGA targets. 
The $2^{\prime\prime}$ fiber diameter corresponds to a $\sim$ 1kpc spatial resolution at the peak redshift ($z \sim$ 0.03) of the MaNGA sample.

\subsection{Data Analysis}\label{sec:dap}
The MaNGA sample and data products used here were drawn from the internal MaNGA Product Launch-6 
(MPL-6), which includes $\sim$4633 galaxies observed through July 2016 (the first three years of the survey). 
The MaNGA data analysis pipeline (DAP, Westfall et al. 2018), which uses pPXF \citep{cappellari04} and a subset of stellar templates 
drawn from the MILES library \citep{sanchez06}, fits the stellar continuum in each spaxel and produces estimates of the stellar absorption lines, alongside measurements of 21 major nebular emission lines in the MaNGA wavelength coverage. 
In addition to analysing each individual spaxel, the DAP also builds spatially Voronoi-binned datacubes using the algorithm of \citet{cappellari03}, and 
performs the analysis on these binned spectra. 

For this study we extract from the DAP products named ``SPX-GAU-MILESHC'' (analysis of each individual pixel) the projected stellar rotation velocity ($v_{\rm star}$), stellar velocity dispersion ($\sigma_{\rm star}$), rotation velocity of ionized gas ($v_{\rm gas}$) and velocity dispersion of ionized gas ($\sigma_{\rm gas}$). From the Voronoi-binned datacubes named ``MAPS-VOR10-GAU-MILESHC'', which are binned to S/N$\sim$ 10 in $r-$band, we extract the spectral indices D$_n$4000 and H$\delta_{\rm A}$,  and nebular emission line fluxes and equivalent widths. Note that emission line fluxes and equivalent width values are corrected for underlying stellar continuum absorption. The index D$_n$4000 measures the strength of the 4000\AA\ break parameterized as the ratio of the flux density 
between two narrow continuum bands 3850$\sim$3950 and 4000$\sim$4100\AA\ \citep{bruzual83}. The Lick Index H$\delta_{\rm A}$ is the equivalent width of H$\delta$ absorption feature in the bandpass 4083$-$4122\AA\ with continuum bandpasses of 4041.6$-$4079.75\AA\ and 4128.5$-$4161.0\AA\ \citep{worthey94, worthey97}.

The derived galaxy parameters required in this work including total stellar mass ($M_*$), lighted-weighted and mass-weighted stellar age, and S$\acute{\rm e}$rsic index. The total stellar mass was taken from the MPA-JHU catalog\footnote{\url{ https://wwwmpa.mpa-garching.mpg.de/SDSS/DR7/}}. The stellar $M/L$ are obtained by comparing $u,g,r,i,z$ colors of galaxies to a large grid of model galaxy colors following the methodology described in \citet{brinchmann04} and \citet{tremonti04}. The light and mass-weighted stellar ages are from MaNGA-Pipe3D value-added catalog \citep{sanchez16a, sanchez16b}. The S$\acute{\rm e}$rsic index are taken from MaNGA PyMorph photometric catalogue \citep{fischer19}.

\subsection{Sample Selection}\label{sec:sample}
\begin{figure}
\bc
\hspace{-0.0cm}
\resizebox{8.5cm}{!}{\includegraphics{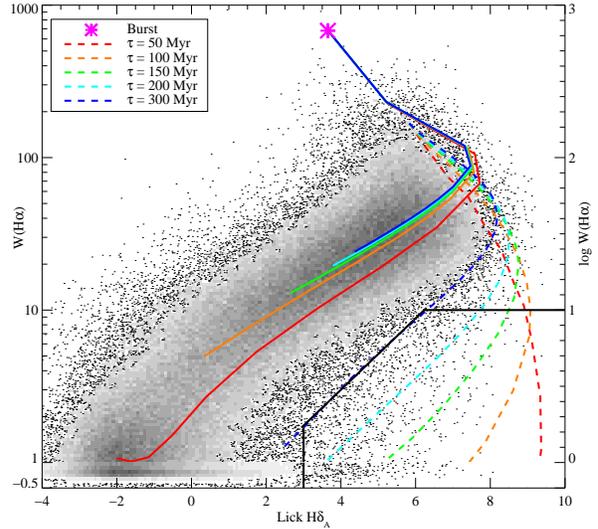}}\\%
\caption{The \hda\ absorption line vs. \ha\ emission line equivalent width for SDSS DR7 galaxies (greyscale) and toy model evolutionary tracks (coloured lines). The solid lines show exponentially declining star formation histories with e-folding times of $\tau=$0.5 Gyr (red) to $\tau=$5 Gyr (blue). The dashed lines have an additional burst of star formation after 6.5 Gyr, followed by truncation with $e$-folding times given
in the legend. For all models the youngest point is marked by the magenta star, and they evolve to the bottom left over time. The black solid line indicates the region of our PSB selection box. Y-axis is in log-space when \ha\ emission line equivalent larger than 1 and in linear space when it is smaller than 1.}
\label{sample}
\ec
\end{figure}

There are multiple ways in which to select PSB galaxies. As described in Section \ref{sec:intro}, traditionally PSBs are identified based on the presence of strong Balmer absorption (an intermediate age stellar population) and weak or
absent emission lines (e.g. H$\alpha$ and/or \oii) indicating no ongoing star formation. Recent studies have challenged the strict limit placed on nebular emission lines, which biases the PSB selection against galaxies hosting narrow line AGN or shocks, and excludes galaxies that are post-starburst but not (yet) fully quenched \citep{yan06, wild07,  wild09, kocevski11,  yesuf14, alatalo14}. However, not selecting on emission lines means some PSBs are missed where their Balmer lines are indistinguishable from the star-forming population, and can  
lead to contamination from dust-obscured starbursts \citep{poggiantiwu00,wild07}.  

In this work, we adjust the traditional method to identify galaxies
with lower H$\alpha$ nebular emission than expected for the strength
of their Balmer absorption lines. This allows us to include galaxies
in the process of shutting down their star formation, although may
still exclude spaxels containing gas excited by shocks or a central
AGN. Fig.~\ref{sample} shows the strong correlation between \Wha\ and
\hda\ for a sample of galaxies from the SDSS Data Release 7 using central fibre spectroscopy. The SDSS-DR7 
sample includes 192,678 galaxies from the MPA-JHU catalog with $z$Warning $=$ 0 and spectral median S/N$>10$ per pixel, redshift in 
the range of $0.01<z<0.08$, and stellar mass in the range of $10^{8}<M*/M_{\odot}<10^{12}$. Overplotted are toy model evolutionary tracks created from \citet{bruzual03} spectral synthesis models assuming a Chabrier \citep{chabrier03} initial mass function (IMF),
where the \ha\ emission is computed from the model spectrum  Lyman continuum flux following appendix B in \citet{hunter04}. The models have an exponentially declining star formation rate with an e-folding time of $\tau \sim$ 5\,Gyr, then a second exponentially declining starburst is added after 6.5\,Gyr of evolution. The tracks for bursts with a range of decline times are shown. We identify PSB spaxels where they lie below the starburst track with an e-folding time of 300\,Myr (marked by the black solid line), the same decline time found by \citet{wild10} for central starbursts in the SDSS DR7. For this model, the \Wha\ falls to 10\AA\ after $\sim1$\,Gyr. 

In summary, we first require the spaxels to have a median spectral S/N $>$ 10 per pixel, in order to obtain a robust measurement of \hda. We then select a spaxel to be a PSB if it has \hda$>$3\AA, \Wha$<$10\AA\ and $\log$\Wha$<0.23\times$\hda$- 0.46$. Clearly the toy models should be taken as indicative only, with the previous star formation history, burst mass fraction and dust content of the galaxy playing a role in the true evolution of these spectral measurements. Our aim here is to select starburst regions that have recently and rapidly truncated their star formation, but equally to include starburst regions that have not yet fully quenched their star formation. 

\begin{figure*}
\bc
\hspace{-0.6cm}
\resizebox{16cm}{!}{\includegraphics{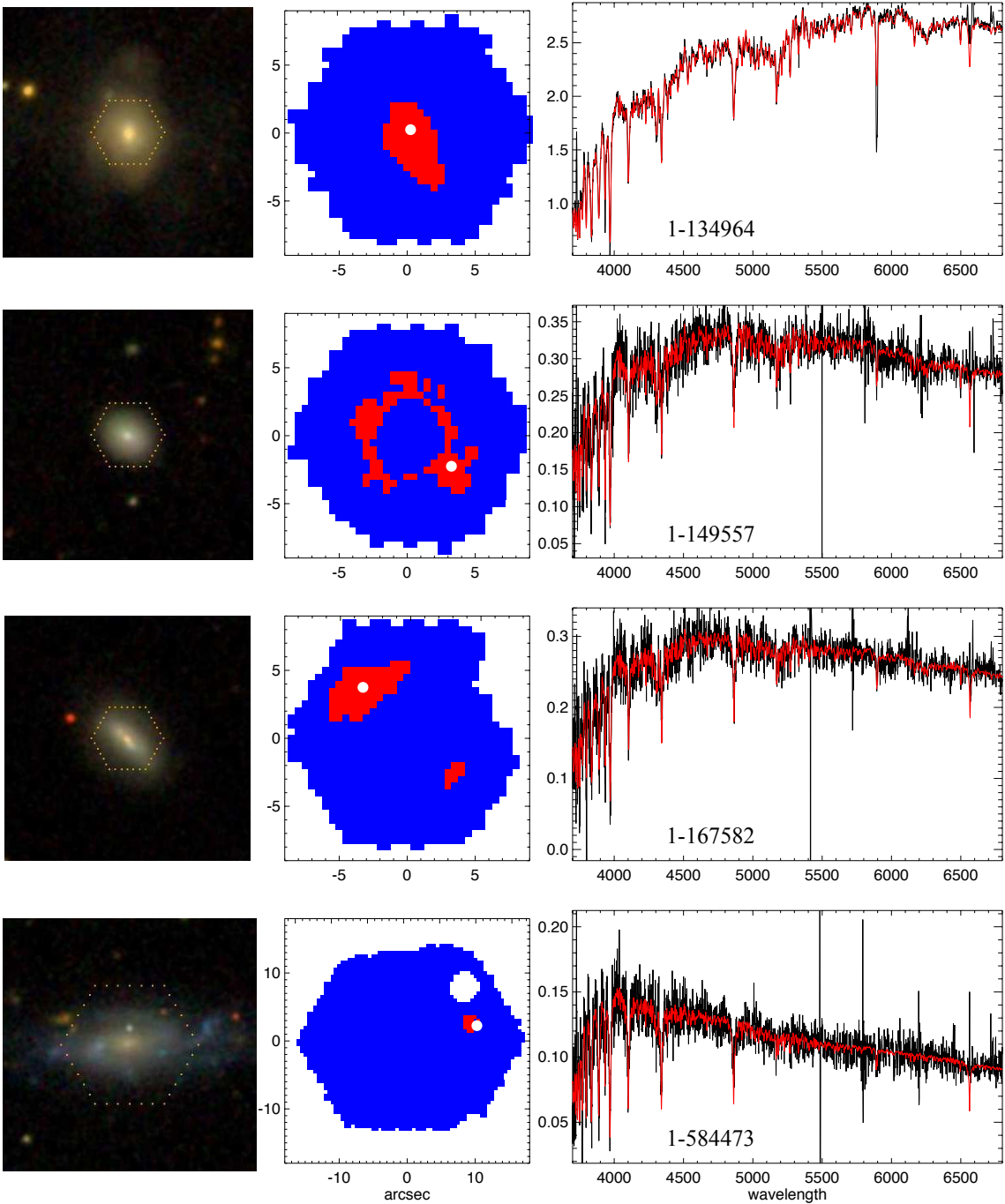}}\\%
\caption{Examples of MaNGA galaxies with PSB regions, MaNGA ID for each galaxy is shown in the right panel. The top row shows a galaxy with central post-starburst regions (CPSB). 
The second and third rows show galaxies with ring-like post-starburst regions (RPSB) and the 
bottom row shows a galaxy with an irregular region in the outskirts (IPSBs). For all four examples, 
the left panel shows the SDSS $g,r,i-$image, the middle panel shows the post-starburst spaxels in red, the spectrum located at the position of the white-solid dot is shown in the right panel, with the observed spectrum in black and the best-fit continuum model 
output by the MaNGA DAP in red. 
\label{sampexp}}
\ec
\end{figure*}

We identify 406/4633 galaxies with more than 6 contiguous spaxels that satisfy these selection criteria. Finally, a careful visual check was done to remove contaminants with foreground stars and background galaxies which affect the continuum fit. This reduces the sample to 360 galaxies. 
Figure~\ref{sampexp} shows four examples of galaxies with PSB regions. The left panel shows the SDSS $g,r,i-$images, 
the middle panel shows the PSB regions in red, the spectrum located  at the position of the white-solid dot is shown in the right panel, where the observed spectrum is in black and the best-fit continuum models output by the MaNGA DAP in red. 

We visually classify these 360 galaxies into three types: galaxies with central post-starburst regions (CPSB, top
row of Figure~\ref{sampexp}); galaxies with ring-like post-starburst regions (RPSB, second and third rows of Figure~\ref{sampexp}); and those with irregular outskirt PSB spaxels (IPSBs, bottom row of Figure~\ref{sampexp}). 
The third galaxy is classified as RPSB since it is edge-on, and the two PSB regions in these RPSBs have similar rotation velocity with inverse rotation direction. In summary, we find 31 CPBs, 37 RPSBs and 292 IPSBs. 
We note that both the RPSBs and IPSBs are entirely new classes of PSBs. 

The 31 CPSBs corresponds to a fraction of 0.7\% (31/4633) over the whole MaNGA sample. This low fraction is consistent 
with the results from \citet{goto03a} and \citet{goto07a}, who claim that the CPSB galaxies are a rare population (less than 1\%) in the local Universe, although the selection criteria is not exactly the same. The incidence of localised PSB regions is clearly a much more common phenomenon. These could be due to individual bright star clusters or cluster complexes with
intermediate ages. We will leave the study of these objects to future work, and here focus on the CPSBs and RPSBs. 

\subsection{Control samples}

In order to quantify the difference between the PSB galaxies and
typical galaxies, we built a control sample of galaxies without PSB
regions. For each PSB, we select 10 control galaxies which are closely
matched in stellar mass and global \d4n. The global spectral indices
are not produced as part of the DAP, therefore we stack the
spectra across the full spatial extent of the MaNGA observations in
order to measure them. 

Our aim is to compare the PSB galaxies to plausible progenitors, therefore the motivation for choosing these two matching parameters is the
following: (i) constraining the control galaxies to have similar
stellar mass is extremely important because stellar population
properties are known to vary strongly with stellar mass, (ii)
constraining global \d4n\ ensures the control samples have
stellar populations with similar light-weighted ages when averaged
over the past few Gyr i.e. prior to any event that caused the post-starburst features. This match will not be perfect, as \d4n\ increases following a 
shut down in star formation, but avoids the need for extensive model dependent spectral fitting that is beyond the scope of this paper. 

\section{Results}

\begin{figure*}
\bc
\hspace{-0.0cm}
\resizebox{17cm}{!}{\includegraphics{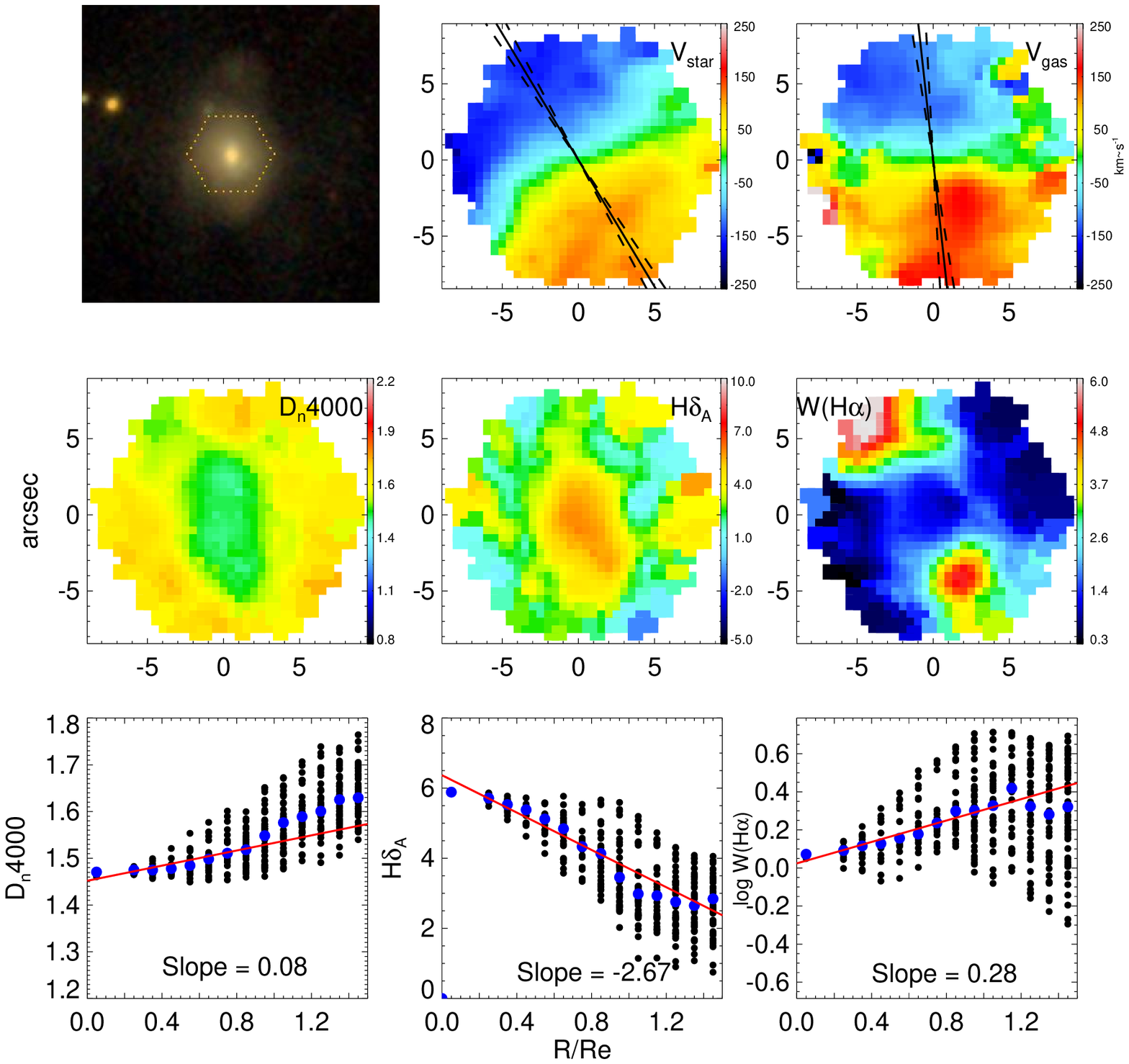}}\\%
\caption{The spatially resolved stellar velocity, gas velocity,
 \d4n, \hda and \Wha\ of an example CPSB galaxy, MaNGA ID: 1-134964. \emph{Top:} the SDSS three-color image, stellar and gas
 velocity fields. The solid line over-plotted on each velocity field
 shows the kinematic position angle, with
the two dashed lines showing the $1\sigma$ error range. \emph{Middle:}
the \d4n, \hda and \Wha maps. \emph{Bottom:}
the radial profiles of \d4n, \hda\ and \Wha. The black dots show the parameter value of each spaxel, blue dots show 
the median values in bins of 0.1 $R_{\rm e}$. The red solid line is a
linear fit to the binned median values, with the value of the fitted slope given in each
panel.
\label{fig:cpsb}}
\ec
\end{figure*}

\begin{figure*}
\bc
\hspace{-0.0cm}
\resizebox{17cm}{!}{\includegraphics{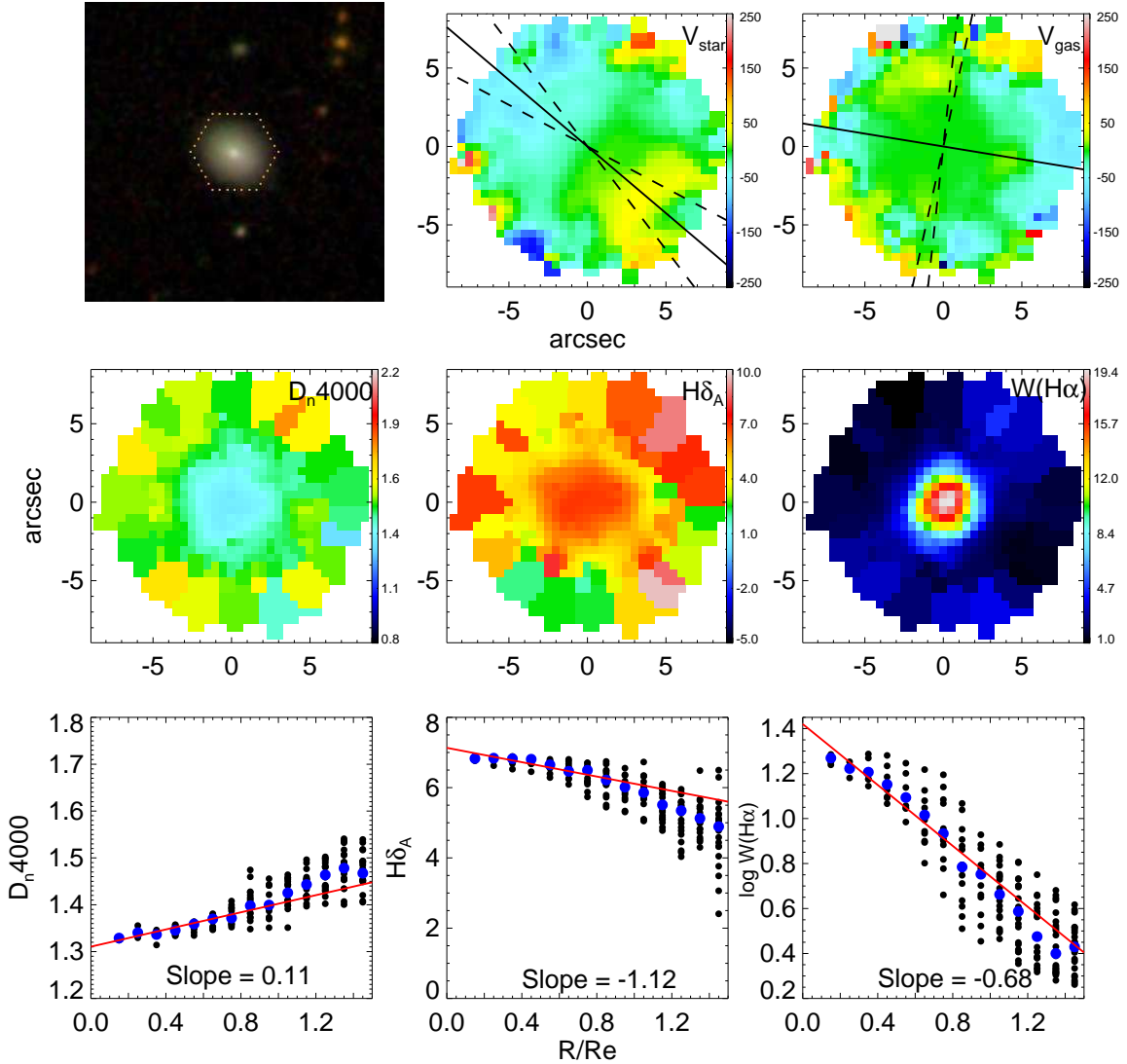}}\\%
\caption{The same as Fig.~\ref{fig:cpsb} for an example RPSB galaxy, MaNGA ID: 1-149557.}
\label{fig:rpsb}
\ec
\end{figure*}

\begin{table*}
\begin{center}
\caption{The sample of CPSB galaxies. (1) MaNGA identifier; (2) Right Ascension; (3) Declination; (4) log $M_*$; (5) \d4n; (6) inclination for galaxies with low bulge fractions (fracDEV $<$ 0.8); (7) S$\acute{\rm e}$rsic index; (8) slope of \d4n gradient; (9) slope of \hda\ gradient; (10) slope of log\Wha\ gradient; (11) kinematic misalignment between gas and stars ($ |\rm{PA_{kin,star}}- \rm{PA_{kin,gas}}|$); (12) notes of unusual features (EML = emission line).}
\centering
\begin{tabular}{lllllllllllp{2cm}}
\hline\hline
MaNGAID & RA & DEC & log$M_*$ & \d4n & $i$ & $n$ & $\nabla_{{\rm D_n4000}}$ & $\nabla_{\rm H\delta_A}$ & $\nabla_{\rm log W(H\alpha)}$ & $\rm{\Delta PA_{kin}}$ & $\rm{Features}$ \\
{(1)} & {(2)} & {(3)} & {(4)} & {(5)} & {(6)} & {(7)} & {(8)} & {(9)} & {(10)} & {(11)} & {(12)}\\
\hline
1-134964     &       246.76069       &       43.476100       &    10.95   &    1.54  &   -      &      3.2    &     0.100 &  -2.963 &   0.286 &   25 $\pm$ 4      &   tidal tail \\
1-146344     &       120.06709       &       29.471437       &      9.97   &    1.49  &   -      &      2.0    &     0.099 &  -1.900 &   0.143 & 180 $\pm$ 26   &   misaligned \\
1-149589     &       170.72345       &       51.341778       &      9.85   &    1.55  &   53   &      1.2    &     0.055 &  -1.112 &  -0.182  &    0 $\pm$ 20    &   - \\
1-149709     &       173.41287       &       52.674587       &    10.19   &    1.47  &   -      &      5.1    &     0.104 &  -3.969 &  -0.283 & 137 $\pm$ 7    &   misaligned  \\
1-152474     &        114.45795      &       28.652892       &      8.85   &    1.53  &   3      &      1.6    &    -0.054 &  -0.437 &   0.244 &     -               &   no EML  \\
1-163965     &       120.05960       &       26.698015       &    10.32   &    1.67  &   -       &      4.5    &     0.133 &  -2.114 &   0.037 & 130 $\pm$ 7    &   misaligned   \\
1-178374     &       260.61132       &        28.309696      &    10.78   &    1.60  &    -       &      8.0    &    0.055 &  -0.991 &  -0.222 &             -               &   tidal tail  \\
1-178823     &       311.76380       &       0.43677787     &      9.40   &    1.60  &   53    &      -    &     0.473 &  -3.173 &  -0.447 &  150 $\pm$ 10   &   misaligned  \\
1-179682     &       317.42261       &       0.62776940     &      9.45   &    1.57  &   -       &      7.2    &     0.079 &  -2.818 &  -0.191 &     -                        &   no projected rotation  \\
1-210114     &       242.58533       &       41.854895       &    10.74   &    1.57  &   -       &      8.0    &     0.042 &  -1.604 &  -0.529 &     -                        &   tidal tail \\
1-248389     &       240.65805       &       41.293427       &    10.59   &    1.62  &   -       &      7.0    &     0.088 &  -2.318 &  -0.977 &   50 $\pm$ 4       &   tidal tail  \\
1-250969     &       206.29627       &       42.319513       &    10.11   &    1.50  &   61    &      1.9    &     -0.047 &  -1.100 &   0.119 &     0 $\pm$ 11       &   bar  \\
1-295343     &       246.48074       &       25.411607       &      9.91   &    1.57  &   -       &      5.0    &      0.099 &  -2.730 &   0.658 &  180 $\pm$ 14    &   misaligned  \\
1-29809       &       358.46882       &      -0.0987309       &      9.45   &    1.59  &   43    &      1.5    &      0.029 &  -1.901 &  -0.317 &  143 $\pm$ 33   &   misaligned \\
1-301834     &      148.42110        &      35.701876         &     9.77   &     1.50  &   -      &      3.3    &     -0.007 &  -0.455 &  -0.057 &   93 $\pm$ 16     &   misaligned \\
1-38062       &       49.228867       &      -0.04200700     &    10.07   &    1.48  &   -       &      4.8    &      0.080 &  -2.955 &  -0.220 &    25 $\pm$ 9      &   disturbed gas velocity field\\
1-38166       &       49.946854       &      0.62382219       &     9.28   &    1.42   &  -       &      4.5    &      0.298 &  -2.349 &  -0.300 &    -                       &   no projected rotation \\
1-38374       &       50.888599       &      -0.43853564      &     9.88   &    1.59   &  69    &      1.5    &     -0.012 &  -1.018 &   0.228 &    93 $\pm$ 7        &   misaligned\\
1-384400     &       126.75586       &        21.706752       &     9.99   &    1.48   &  78    &      1.8    &      0.074 &  -2.367 &  -0.688 &     90$\pm$ 4       &   misaligned     \\
1-384486     &       127.31796       &        23.809021       &     9.28   &    1.52   &   -       &      6.5    &     0.153 &  -6.250 &   0.699 &     -                       &    no EML\\
1-385499     &       129.99929       &        23.413400       &     8.97   &    1.32   &   70    &      0.7    &    -0.025 &   0.487 &   0.136 &     0$\pm$11         &    -           \\
1-404249     &       194.52342       &        29.017353       &     9.37   &    1.50   &   82    &      1.8    &    -0.048 &  -0.256 &   0.147 &    -                        &    no EML \\
1-43584       &       117.06113        &        39.045731       &     8.91   &    1.26   &   81    &      0.7    &    -0.099 &   1.137 &   0.545 &    12$\pm$62       &    -            \\
1-44447       &       120.63984       &        42.392705       &    10.04  &    1.56    &  -       &       3.5    &    0.096 &  -2.585 &  -0.365 &    18$\pm$63        &   disturbed gas velocity field \\
1-456744     &       194.33162       &        27.613856       &      9.16  &     1.54   &  74    &       1.7    &    0.114 &  -0.447 &   0.460 &    -                        &    no EML \\
1-456850     &       194.63449       &        28.377961       &      8.62  &     1.42   &  44    &       1.3    &    0.204 &   2.281 &   0.885 &   -                        &     no EML\\
1-457004     &       196.26374       &         27.537037      &      9.17  &     1.49   &  64    &       1.7     &  -0.045 &   0.507 &   0.797 &      -                        &   no EML\\
1-457130     &       195.33050       &         27.860463      &      9.02  &     1.35   &  3      &       2.0     &   0.174 &  -1.345 &  -0.842 &     -                        &   no projected gas rotation\\
1-560826     &       236.16573       &        38.425357       &    11.03  &     1.61   &  32    &       5.2    &    0.192 &  -3.816 &  -0.448 &      -                        &    tidal tail\\
1-72913       &       127.48937       &        44.940158       &    10.79  &     1.62   &  -       &       4.4    &    0.045 &  -1.892 &  -0.318 &     124$\pm$37      &   misaligned\\
12-98126     &       230.50740       &        43.534632       &      9.69  &     1.59   &  43    &       5.7     &  -0.075 &   2.678 &   0.140 &     -                        &   no EML\\
\hline
\end{tabular}\\
\end{center}
\label{tab:cpsb}
\end{table*}

\begin{table*}
\begin{center}
\caption{The sample of RPSB galaxies. (1) MaNGA identifier; (2) Right Ascension; (3) Declination; (4) log $M_*$; (5) \d4n; (6) inclination  for galaxies with low bulge fractions (fracDEV $<$ 0.8); (7) S$\acute{\rm e}$rsic index; (8) slope of \d4n gradient; (9) slope of \hda\ gradient; (10) slope of log\Wha\ gradient; (11) kinematic misalignment between gas and stars ($ |\rm{PA_{kin,star}}- \rm{PA_{kin,gas}}|$); (12) notes of unusual features (EML = emission line).}
\centering
\begin{tabular}{lllllllllllp{2cm}}
\hline\hline
MaNGAID & RA & DEC & log$M_*$ & \d4n & $i$ & $n$ & $\nabla_{{\rm D_n4000}}$ & $\nabla_{\rm H\delta_A}$  & $\nabla_{\rm log W(H\alpha)}$ & $\rm{\Delta PA_{kin}}$ & $\rm{Features}$ \\
{(1)} & {(2)} & {(3)} & {(4)} & {(5)} & {(6)} & {(7)} & {(8)} & {(9)} & {(10)} & {(11)} & {(12)}\\
\hline
1-134004        &       238.44858       &      47.404958        &   9.38 &    1.48 &       57   &       1.4    &  0.136 &  -0.978 &  -0.548 &       149 $\pm$ 22        &        misaligned \\
1-149557        &       171.77902       &       51.131645       &   9.15 &    1.42 &       38   &       1.4    &  0.102 &  -1.137 &  -0.703 &       156 $\pm$ 87        &        migalign \\
1-153247        &       119.36531       &       33.257935       &   8.94 &    1.44 &       59   &       1.7    &  0.088 &  -2.071 &  -0.703 &        149 $\pm$ 90       &        misaligned \\
1-167582        &       154.48083       &       46.603286       &   9.94 &    1.37 &       69   &       1.5    &  0.122 &  -1.391 &  -0.994 &       43 $\pm$ 18          &        misaligned  \\
1-201355        &       117.05387       &       28.225092       &  10.14 &    1.30 &         3   &       -    &  0.073 &   0.789 &  -0.441 &        6 $\pm$ 8             &        interacting/ pre-merger \\
1-211002        &        247.14945      &       39.719740       &   9.70 &    1.49 &        37  &        0.7   &  0.090 &   0.044 &  -0.544 &        32 $\pm$ 16         &        misaligned\\
1-216976        &        135.75897      &      40.433985       &  10.60 &    1.45 &         6    &       4.5    &  0.257 &  -3.853 &  -1.487 &       -                             &        tidal tail\\
1-217015        &       136.11419       &       41.486207       &   9.18 &    1.36 &       51   &        1.8    &  0.053 &  -0.065 &  -0.249 &       6 $\pm$ 17           &         tidal tail \\
1-217221        &       138.75315       &       42.024390       &  10.27 &    1.23 &       64   &       1.0     &  0.104 &   0.874 &  -0.572 &       0 $\pm$ 4             &        disturbed gas velocity field  \\
1-258306        &       183.57898       &       43.535279       &   9.54 &    1.24 &       72   &       2.1     &  0.150 &   0.552 &  -0.960 &       12 $\pm$ 11         &         disturbed gas velocity field \\
1-258380        &       181.54597       &       45.149206       &  10.96 &    1.43 &       -      &       4.0     &  0.031 &  -0.464 &  -0.477 &       0 $\pm$ 4             &         - \\
1-277246        &       166.18780       &       45.156430       &   9.37 &    1.52 &       20   &       1.6     &  0.246 &  -2.150 &  -0.937 &       -                            &         face-on, no projected rotation\\
1-277691        &       164.58519       &       40.788234       &   9.63 &    1.34 &       76   &       0.7     &  0.002 &  -0.005 &  -0.031 &       0 $\pm$ 9             &          edge-on\\
1-29512          &       356.75183       &       -0.4473874      &  11.13 &    1.34 &       -      &       4.2     &  0.163 &   1.470 &  -0.471 &       0 $\pm$ 7             &          bar\\
1-321354        &       218.94756       &       47.007467       &   9.59 &    1.31 &       66   &       1.6         &  0.134 &  -1.272 &  -0.264 &       6 $\pm$ 15           &          edge-on \\
1-373878        &       228.41485       &      28.244461       &   9.94 &    1.28 &       87   &       1.1     &  0.083 &   0.403 &  -0.814 &       0 $\pm$ 13            &          interacting/ pre-merger\\
1-37862          &       47.029452       &       0.4562083       &  10.96 &    1.41 &       -      &       6.0     &  0.066 &  -0.250 &  -0.399 &        12 $\pm$ 4           &          tidal tail\\
1-38041          &       49.457454       &      -0.5546585       &   9.94 &    1.52 &       66    &       1.4     &  0.187 &  -3.188 &  -1.609 &       124 $\pm$ 10        &          misaligned \\
1-38168          &       49.929339       &       0.5654778       &  10.06 &    1.28 &       -      &       5.6     &  0.182 &   0.765 &  -1.115 &       0 $\pm$ 11            &          interacting/ pre-merger   \\
1-38470          &       51.708914       &      0.19858883       &   9.74 &    1.34 &      -      &       5.3     &  0.115 &  -0.722 &  -0.032 &       12 $\pm$ 18          &          tidal tail\\
1-386695        &       137.98351       &       27.899270       &  10.11 &    1.31 &       76   &       1.3     &   0.152 &   1.440 &  -0.686 &      7 $\pm$ 11             &          bar     \\
1-387081        &       139.17787       &       28.054233       &  10.27 &    1.41 &       86   &       1.0     &  0.095 &  -0.594 &  -0.877 &        18 $\pm$ 16           &         edge-on \\
1-392007        &       154.97835       &       36.325739       &  10.20 &    1.49 &      72     &       2.4     & -0.021 &  -0.003 &  -0.202 &       6 $\pm$ 4               &         edge-on\\
1-405760        &       196.10272       &       36.479950       &  10.30 &    1.48 &       87   &       1.0     & -0.048 &  -0.394 &  -0.751 &       178 $\pm$ 4            &         misaligned     \\
1-419380        &       183.00790       &       35.404399       &    9.89 &    1.41 &       57   &       3.6     &  0.290 &  -3.467 &  -1.290 &       172 $\pm$ 6            &         misaligned     \\ 
1-456309        &       194.76938       &       26.958192       &    9.42 &    1.47 &       73   &       4.1     &   0.013 &   2.594 &   0.262 &      -                               &         no EML\\
1-456915        &       194.73314       &       27.833445       &  10.54 &    1.51 &       87   &       0.8     &  -0.164 &   1.821 &  -0.043 &      18 $\pm$ 4              &         disturbed gas velocity field     \\
1-457200        &       196.47287       &       28.112434       &  10.29 &    1.43 &       71   &       1.7     &  0.036 &  -0.200 &  -1.348 &       18 $\pm$ 12            &         bar\\
1-548626        &       120.55787       &       37.150076       &  10.66 &    1.40 &       87   &       1.4     & -0.152 &   2.877 &  -0.627 &        0 $\pm$ 4               &         interacting/ pre-merger\\
1-558926        &       140.41142       &       43.726152       &  10.38 &    1.41 &       66   &       1.2     &  0.057 &  -0.942 &  -0.409 &       0 $\pm$ 4                &         tidal tail     \\
1-574504        &       123.82033       &       46.075253       &  10.50 &    1.26 &       62   &       -     &  0.415 &   4.120 &  -1.183 &       12 $\pm$ 4              &         bar     \\
1-585632        &       143.51035       &       50.027486       &  10.45 &    1.54 &       23   &       2.4     & -0.007 &   0.835 &  -0.946 &       -                               &         interacting/ pre-merger \\
1-606105        &       147.66431       &       44.331163       &    9.48 &    1.47 &      60   &       1.9     &  0.114 &  -2.462 &  -1.336 &       31 $\pm$ 40            &         misaligned     \\
1-625070        &       198.78425       &       30.403775       &  10.35 &    1.37 &       43   &       1.0     &  0.149 &   1.232 &  -0.984 &       12 $\pm$ 4              &         bar     \\
1-626502        &       203.05706       &       26.949981       &  10.52 &    1.48 &       74   &       1.2     &  0.058 &   0.163 &  -1.133 &       0 $\pm$ 4                &         bar     \\
1-630590        &       218.97634       &       53.391637       &  10.36 &    1.53 &       75   &       1.0     &  0.075 &  -0.129 &  -1.061 &       0 $\pm$ 6                &         bar\\
1-633000        &       233.23196       &       42.438257       &  10.10 &    1.46 &       64   &       2.0     &  0.118 &  -1.098 &  -0.999 &       156$\pm$ 18           &         misaligned\\
\hline
\end{tabular}\\
 \end{center}
\label{tab:rpsb}
\end{table*}

Tables 1 \& 2 list the samples of CPSB and RPSB galaxies, together with relevant parameters used in this paper. Figures \ref{fig:cpsb} and \ref{fig:rpsb} show examples of the 
spatially resolved stellar velocity, gas velocity, \d4n, \hda\ and
\Wha\ maps of an example CPSB and RPSB galaxy that are broadly typical
of the samples. The bottom panels show the radial gradients of the
spectral indices, in units of the
effective radius. 

In the following sections we discuss the global
properties of the CPSB and RPSB host galaxies, their 
kinematics and stellar populations. We are interested in understanding
whether the new RPSB population is related to the CPSB population, and
whether the spatially resolved maps are consistent with the hypothesis
that (C)PSB galaxies are caused by major mergers.  



\subsection{Global population properties}

\begin{figure*}
\bc
\hspace{-0.6cm}
\resizebox{17cm}{!}{\includegraphics{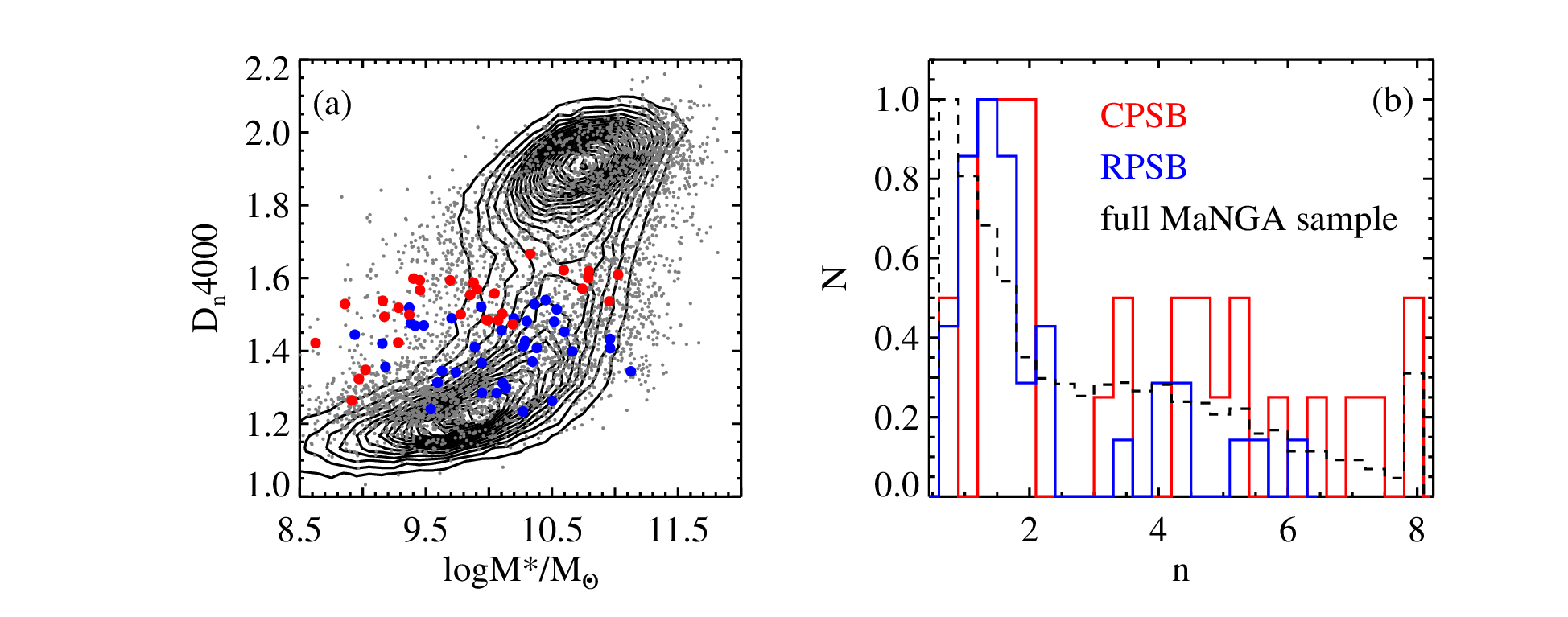}}\\%
\caption{\emph{Left:} The global \d4n--stellar mass relation for the CPSB and
  RPSB galaxies (red and blue dots respectively), with the SDSS DR7 sample as
  contours and full MaNGA sample as grey dots. 
\emph{Right:} The S$\acute{\rm e}$rsic index distribution for the RPSB and 
CPSB galaxies as blue and red histograms, respectively. The black dashed-line shows the full MaNGA sample.
}\label{fig:global}
\ec
\end{figure*}

The left panel of Figure \ref{fig:global} shows the CPSB and RPSBs on
the global \d4n- stellar mass relation as red and blue dots
respectively. Contours show the SDSS DR7 sample (as described in
Section \ref{sec:sample}), while the grey dots show the full MaNGA
sample. For the MaNGA galaxies, \d4n\ is measured from the global
spectrum, whereas for the SDSS DR7 sample, \d4n\ is measured from the
central fibre spectroscopy. We choose to show \d4n\ rather than SFR or
sSFR as it can be measured consistently for all galaxies.  Plotting
global SFR from the MPA-JHU catalogue instead of \d4n\ provides
qualitatively the same picture, but relies on extrapolation of the
fibre measurements using galaxy colours which we found to be biased
for the PSB galaxies in MaNGA.
It is clear that most RPSBs are located on the star forming main sequence 
while the CPSBs are primarily located in the green valley, between the
red and blue sequence.

\citet{fischer19} fit all the MaNGA DR15 galaxies with a single
S$\acute{\rm e}$rsic profile, released as part of the MaNGA PyMorph
catalog.  The right panel of Figure \ref{fig:global} shows the
distribution of S$\acute{\rm e}$rsic index ($n$) for CPSBs (red) and
RPSBs (blue). These values are reproduced in Tables 1 and 2 for the CPSB and RPSB samples respectively. A higher fraction of the CPSB galaxies than RPSB galaxies have $n>3$, consistent with disk-free elliptical galaxies. 
However, both samples cover a wide range of values indicating
both types are hosted by galaxies with diverse morphologies. This is
consistent with results found for CPSBs in SDSS DR7 \citep{pawlik18}. 

\subsection{Stellar populations}
In this section, we study the stellar population distribution of the
galaxies with PSB regions using continuum spectral indices \d4n\ and
\hda\, as well as the H$\alpha$ emission line. 
Figures \ref{fig:cpsb} and \ref{fig:rpsb} show maps and radial
profiles of \d4n, \hda\ and
\Wha\ for the example CPSB and RPSB
galaxies. These galaxies are typical of their classes, and the maps
immediately illustrate the primary differences between them.

\begin{figure*}
\bc
\hspace{-0.0cm}
\resizebox{17cm}{!}{\includegraphics{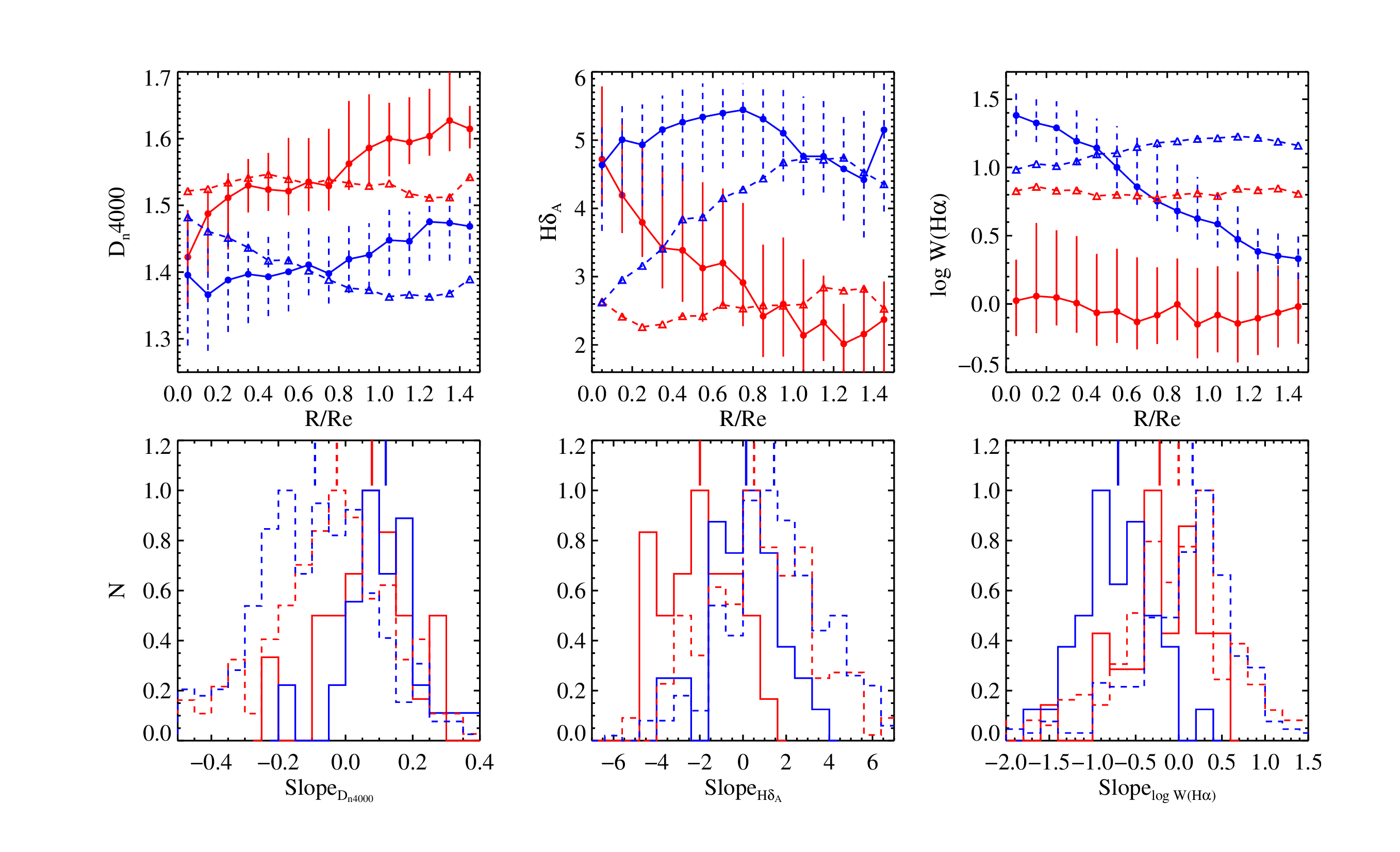}}\\%
\caption{\emph{Top panel:} Radial gradients of \d4n, \hda and \Wha\ for the CPSB
  (red-solid line), RPSB (blue-solid line) as well as their 
  control samples (red-dashed and blue-dashed lines).
  The error bars show the 30\% to 70\% percentile of the distributions for the CPSBs and RPSBs. \emph{Bottom panel:} The distributions of the slope of \d4n, \hda\ and \Wha\ for the four samples, the median values are shown as vertical lines on the top of each panel.}
\label{fig:radgrad}
\ec
\end{figure*}

The top panel of Figure \ref{fig:radgrad} shows the averaged radial profiles of \d4n, \hda\ and \Wha\ for the CPSB (red solid line) and RPSB (blue solid line) samples. The error bars show the 30\% to 70\% percentile of the distributions for the CPSBs and RPSBs. 
The relevant control samples are shown as dashed lines. The bottom panel shows the distributions of the radial profile gradients of \d4n, \hda\ and log\Wha. These slopes are reproduced in Tables 1 and 2. Both the CPSB and RPSB samples have positive
gradients in \d4n, indicating younger stellar populations in the
central regions, with CPSBs having older stellar populations (higher
\d4n) on average than RPSBs across the entire galaxy. However, the
CPSB and RPSB samples differ in their radial profiles of both \hda\
and \Wha: while in the CPSBs the Balmer absorption decreases with
radius and \Wha\ is weak or absent and almost flat, the RPSBs show
strong Balmer absorption over the whole galaxy, while the \Wha\ is
strong in the centre and decreases sharply with radius. We have
verified that the central H$\alpha$ emission in the RPSBs is primarily
contributed by ongoing star formation rather than shocks or AGN: 33 of
37 RPSBs show star-forming/composite-like line ratios on the most
commonly used \citet{bpt} diagram of [NII]/H$\alpha$ vs. [OIII]/H$\beta$ flux ratios  \citep{kewley01, kauffmann03}.

It is the differing radial gradients that lead to the different
classifications of central or ring-like PSB: the CPSBs are typically
only classified as PSBs in the centre, as their Balmer absorption
weakens with radius, whereas the RPSBs are not classified as PSBs in
the centre due to their strong central \Wha. This shows that while the
CPSBs have suppressed star formation throughout their bulge and disk,
and clear evidence of rapid quenching (i.e. strong Balmer absorption) only in the central regions, 
the RPSBs only show clear evidence of recently rapidly suppressed star 
formation in their outer regions. This difference in central star formation also 
explains the different locations of the 2 populations in the global \d4n-stellar mass relation with the RPSBs being located in the star-forming main sequence.

The stellar population gradients for the control samples are also
shown in Figure \ref{fig:radgrad} as dashed lines and triangles.  For
the CPSBs, we find that their \d4n\ profiles are slightly more positive on average than the control sample, consistent with a recent centrally concentrated star-formation episode occuring in the past several Gyr. However, the weak \Wha\ compared to the controls indicates that ongoing star formation has been strongly suppressed across the whole disk. The excess of Balmer absorption in the central regions of the CPSBs further indicates a rapid shut down in star formation has occurred in the last Gyr, which is not seen in the controls. The RPSBs differ from their controls in all three indices, with the controls having a negative gradient in \d4n\ and strong positive gradient in \hda, indicating older stellar populations in the centre, as expected for ordinary bulge-dominated star-forming galaxies.  The controls also show a constant \Wha\ with radius, indicating ongoing star formation throughout the galaxy, significantly different from the sharp decline with radius seen in the RPSBs. We can see that \hda$\sim5$\AA\ is not unusual in the outer regions of ordinary star-forming galaxies; it is only the combination of strong Balmer absorption and weak emission line strength that makes the outer regions of the RPSBs stand out. The \hda$\sim5$\AA\ in the central region of the RPSBs is consistent with ongoing star formation rather than a quenched population (see Figure \ref{sampexp}). Compared to the control samples, this ongoing star formation is significantly stronger, perhaps indicating a time limited starburst rather than more ordinary long-term star formation. 

There are several possible scenarios that could cause the observed
radial gradients in stellar populations of the two samples. It is
possible that either the outer regions of the CPSBs never underwent a
starburst, or that the post-starburst is fading more quickly with
increasing radius, which in turn could imply either that the burst was
weaker in the outer regions, or that it occurred earlier. For the RPSBs, the starburst nature of the central regions combined with the quenching of the outer regions suggests that either a starburst has migrated from the outside inwards, or that the process that caused a central starburst has similarly led to the depletion of gas in the outer regions, potentially through strong gas flows from the outside-in. 



\begin{figure*}
\bc
\hspace{-0.0cm}
\resizebox{17cm}{!}{\includegraphics{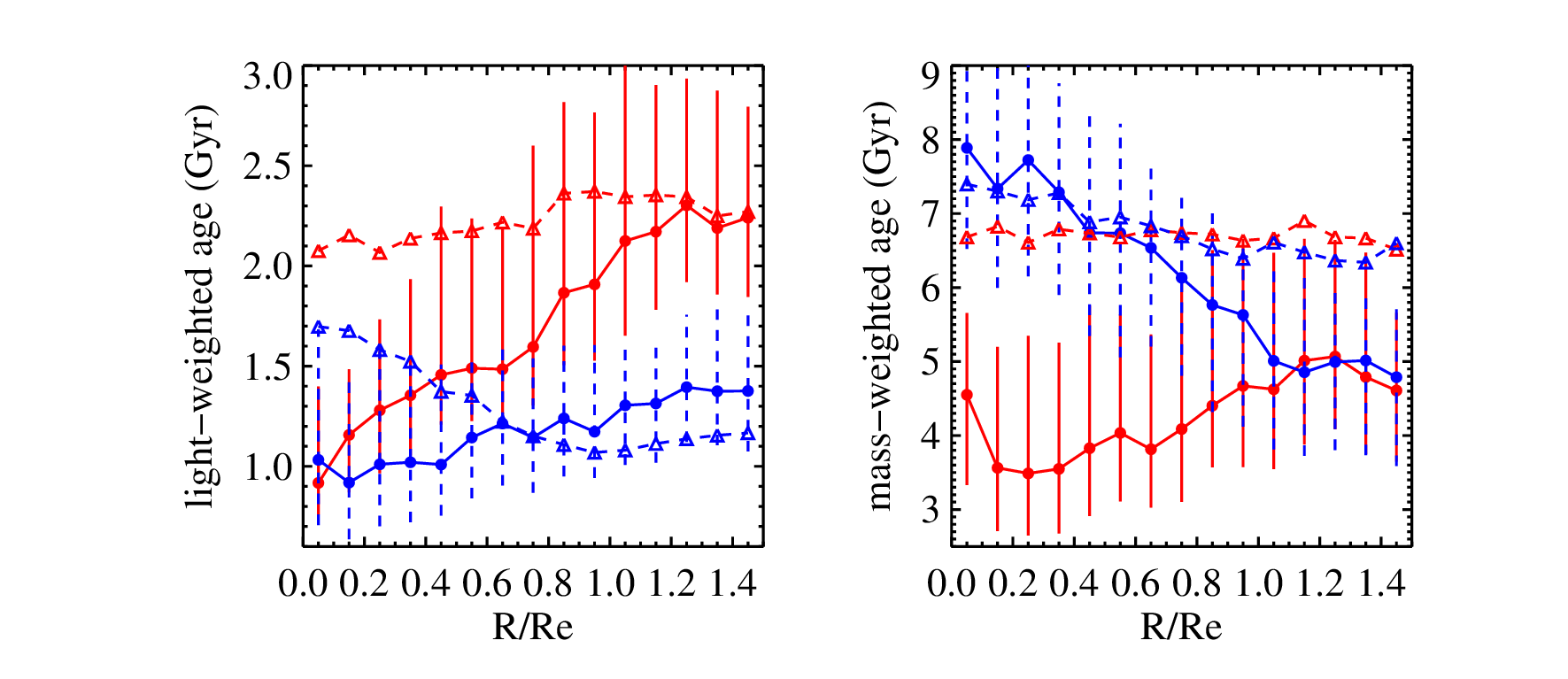}}\\%
\caption{Average lighted-weighted and mass-weighted stellar age of the PSB samples and their controls. The symbols and lines are the same as that in \ref{fig:radgrad}.
\label{fig:age}}
\ec
\end{figure*}

Figure \ref{fig:age} shows the radial gradients in lighted-weighted and mass-weighted stellar age for the PSB samples and their controls,  from the Pipe3D Value Added Catalog \citep{sanchez16a, sanchez16b}. Since \d4n\ is a good indicator of light-weighted age of a stellar population, the radial gradients of light-weighted age and \d4n\ are similar. Light-weighted age enhances the differences between the central regions of the PSBs and the control samples, due to the large fraction of very young stars which dominate the light-weighted age more than they contribute to \d4n. Interestingly, the mass-weighted ages of the RPSB galaxies indicates a substantial old stellar population in their centres (8\,Gyr), consistent with the control sample. This suggests the current central star formation is simply a ``frosting'' on top of a dominant old central population. The mass-weighted age profiles of the CPSBs indicates a very different history, with a higher fraction of stellar mass formed more recently throughout the whole galaxy compared to both control samples and the RPSBs. The very different behaviors in the mass-weighted age of CPSBs and RPSBs indicates that they have very different star formation histories, thus it is impossible for the RPSBs to evolve into CPSBs through secular process. Equally, the similarity in mass-weighted age between the two control samples, contrasted with the difference in the CPSBs supports significant disruption of the stellar component during the event that triggered the post-starburst features. Such a disruption event is not as evident in the RPSB sample. 

\begin{figure*}
\bc
\hspace{-0.0cm}
\resizebox{17cm}{!}{\includegraphics{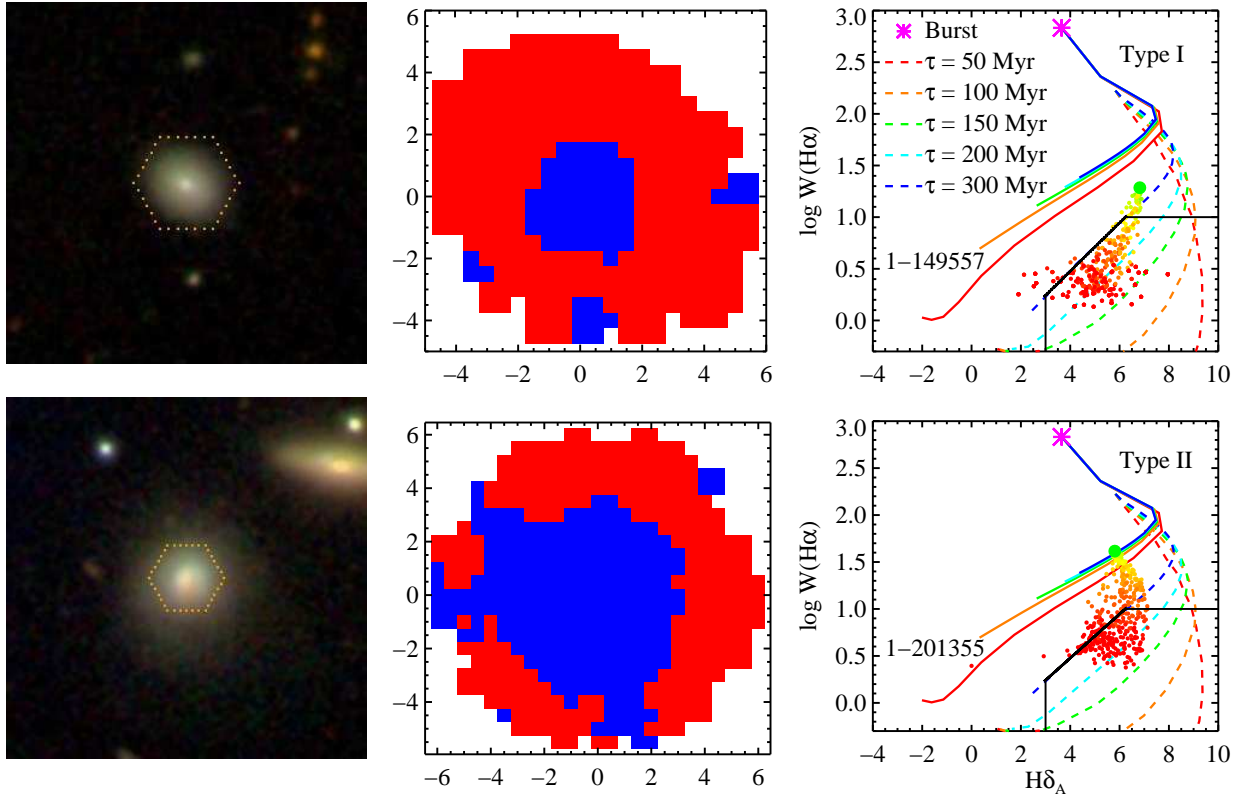}}\\%
\caption{Examples of RPSBs on the \hda\ vs. \Wha\ plane, MaNGA ID of each galaxy is shown in the right panel. We separate RPSBs into type I (top) \& II (bottom). From left to right we show: the SDSS false-color images; the PSB regions in blue; the galaxy spaxels in the  \hda\ vs. \Wha\ plane. The solid and dashed lines are model evolutionary tracks (see Fig.~\ref{sample}). The data points are color-coded by $R/R_{\rm e}$ with the larger green point marking the central pixel and the radius increases as the points become redder. Black lines mark the PSB selection criteria.
\label{fig:rpsbtype}}
\ec
\end{figure*}

Finally,  we investigate the distribution of PSB spaxels in the  \hda\ vs. \Wha\ plane for the RPSB galaxies with data points color-coded by $R/R_{\rm e}$. Figure \ref{fig:rpsbtype} shows examples of two seemingly different classes of RPSB galaxies. At the top (type I) the entire galaxy lies below the star-forming main sequence, with spaxels falling diagonally along the constant-burst-strength model track. At the bottom (type II), the spaxels fall vertically, with the central region still lieing on the star-forming main sequence, while the outer spaxels have lower \Wha. 

The type I RPSBs appear to be undergoing a global shut down in star formation, similar to that seen in the CPSBs but less complete, and is consistent with our constant-burst-strength models with the starburst occurring first in the outer regions of the galaxy and moving inwards over time.  Alternatively, the outer regions could be quenched first, for example due to strangulation processes where the infall of fresh gas is shut off. 

The type II RPSBs are consistent with having a range of quenching timescales, with the outer regions quenching more quickly than the inner regions. It is possible that the inner regions will continue to form stars following the event that caused the outer regions to quench. Alternatively these galaxies are undergoing a rapid quenching following a starburst that occured earlier in the outer regions than the inner regions, similar to the Tpye I RPSBs, but at a later time than the 6.5\,Gyr assumed by our toy models (moving the red dashed line to the left). This type could be caused by ram pressure stripping - the outer regions are most vulnerable to being stripped and quench the quickest. Or they might be linked to mergers or interactions that cause the gas in the galaxies' outskirts to flow inward causing a central starburst and rapid outer quenching. 

Again, the question is whether these two types are evolutionarily linked, with Type II being younger than Type I, or whether they are caused by distinct processes. With more extensive modelling, the current PSB samples may provide a way to disentangle various suggested quenching mechanisms. 


\subsection{Kinematics and morphological features}\label{sec:kin}

\begin{figure}
\bc
\hspace{-0.0cm}
\resizebox{8.5cm}{!}{\includegraphics{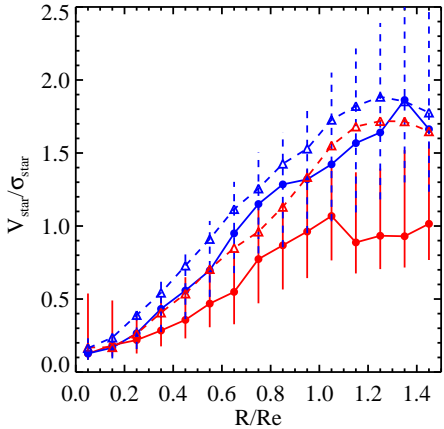}}\\%
\caption{Averaged V$_{\rm star}/\sigma_{\rm star}$ versus the radius for CPSB (red-solid line) and RPSB (blue-solid line) samples,
as well as their control samples (red-dashed and blue dashed lines). The error bars indicate the 30th and 70th percentiles of the distribution.
\label{fig:kine}}
\ec
\end{figure}

In the present-day Universe, the majority of galaxies ($>$85\%) are consistent with being axisymmetric rotating oblate spheroids and only a minor fraction of galaxies have complex dynamics (for a review, see Cappellari 2016). The ratio of ordered to random stellar motion in galaxies has a strong dependence on luminosity or stellar mass (Illingworth 1977; Davies et al. 1983; Emsellem et al. 2011; Brough et al. 2017; van de Sande et al. 2017a; Veale et al. 2017a; Green et al. 2018), which suggests a link between the build-up of stellar mass and angular momentum over time. Major mergers are key candidates leading to a dramatic change in the morphology and spin of galaxies, but ultimately mergers are only one of many physical processes at play, and continuing gas accretion and star formation can reshape the remnant morphology and kinematics (Naab et al. 2014, and citations therein). 

Figure \ref{fig:kine} shows the average stellar velocity to dispersion ratio, $v_{\rm star}/\sigma_{\rm star}$, vs. radius for the CPSB (red solid) and RPSB (blue solid) galaxies, as well as their control samples (red and blue dashed lines). The error bars indicate the 30th and 70th percentiles of the distribution. $v_{\rm star}$ and $\sigma_{\rm star}$ are calculated from the spectral fitting described in Section \ref{sec:dap}. A higher (lower) value of $v_{\rm star}/\sigma_{\rm star}$ corresponds to more (less) rotational support, and therefore this plot can help us to discern any difference in the formation/interaction history of the PSB samples. 
Comparing Figure \ref{fig:kine} with Figure 2 of \citet{emsellem07}, we find that the averaged $v/\sigma$ radial profiles are consistent with fast rotators for all four samples, which is not surprising given the high mass and rareness of slow rotators. However, the CPSBs have noticeably lower $v/\sigma$ than the other three samples, while the RPSB galaxies are much more consistent with their control sample. This indicates that the CPSBs have suffered from more frequent and/or violent interactions, mergers or gas accretion processes \citep{lagos18}. 
 
Such processes can also lead to the inflows of gas required to induce both the excess central star-formation in the RPSB galaxies compared to their controls, and the central post-starbursts in the CPSB galaxies. For example, accretion of counter-rotating gas from a gas-rich dwarf or the cosmic web onto a star-forming galaxy will lead to the redistribution of angular momentum from gas-gas collisions between the pre-existing and the accreted gas, which greatly accelerates gas inflow (Chen et al. 2016). However, additional processes such as bars may play a role \citep{hawarden86, lin17, chown19}. 


In order to understand the possible prevalence of such mechanisms in the PSB samples, we investigate the kinematic misalignment between stars and ionized gas, measured as the difference in the kinematic position angle, $\rm{\Delta PA_{kin} = |PA_*-PA_{gas}|}$, where $\rm{PA_*}$ is the kinematic position angle of the stars and $\rm{PA_{gas}}$ is the kinematic position angle of the ionized gas. The kinematic $\rm{PA}$ is measured using FIT\_KINEMATIC\_PA\footnote{\url{https://www-astro.physics.ox.ac.uk/~mxc/software/}} \citep{krajnovic06}, and is defined as the counter-clockwise angle between north and a line that bisects the velocity field of gas or stars, measured on the receding side.  The solid lines in the top row of Figures \ref{fig:cpsb} and \ref{fig:rpsb} show the best fit position angle to our example galaxies, while the two dashed lines show the $\pm$1$\sigma$ error. The example CPSB galaxy is a star-gas misaligned galaxy with $\rm{\Delta PA_{kin}} = 25$deg, while the example RPSB shows no regular rotation in ionised gas. The values are listed in Tables 1 and 2 for the CPSBs and RPSBs respectively. 
There are 13 CPSBs and 4 RPSBs without $\rm{\Delta PA_{kin}}$ measurements either due to no emission lines or no obvious rotation, we do not consider them in the analysis of misaligned kinematics.

\begin{figure}
\bc
\hspace{-0.0cm}
\resizebox{8.5cm}{!}{\includegraphics{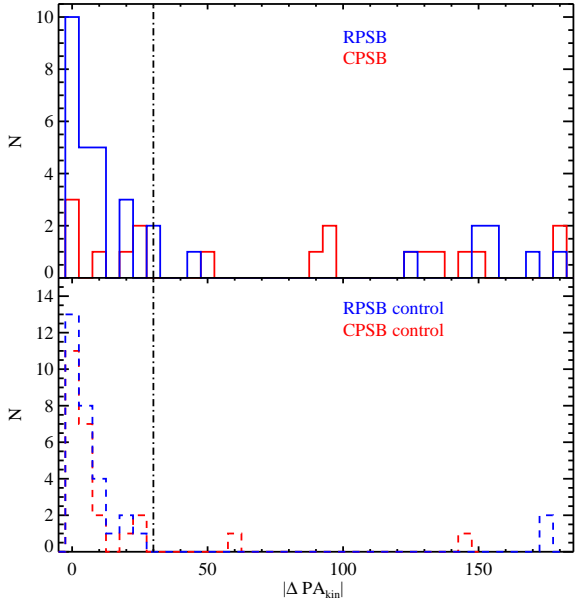}}\\%
\caption{The distribution of $\Delta {\rm PA}_{\rm kin}$ ($= |{\rm PA}_*-{\rm PA}_{\rm gas}|$) for CPSB (red) and RPSB (blue) samples in the top panel and the relevant control samples in the bottom panel. The vertical black lines mark the place where $\Delta {\rm PA}_{\rm kin} = 30$deg, typically used to delineate `normal' from `misaligned' rotation.
\label{fig:pa}}
\ec
\end{figure}

Additionally we use deep images to identify bars, tidal tails and other interaction features in our samples. We use the Legacy Surveys\footnote{\url{http://legacysurvey.org}}, which combines imaging projects on different telescopes, including the Beijing-Arizona Sky Survey (BASS), the DECam Legacy Survey (DECaLS) and the Mayall $z$-band Legacy Survey (MzLS). The images are 1--2 magnitudes deeper than SDSS. Our visual classifications are given in the final column of Tables 1 and 2. 

The top panel of Figure~\ref{fig:pa} shows the distribution of $\rm{\Delta PA_{kin}}$ for CPSB (red) and RPSB (blue) samples, while the $\rm{\Delta PA_{kin}}$ distribution of the control samples are shown in the bottom panel. 
Of the 17 CPSB galaxies for which we have both stellar and emission line kinematic measurements, 10 are misaligned (defined as $\rm{\Delta PA_{kin}} \ge $30 deg). A further 7 show other signs of disturbance such as tidal tails or disturbed gas velocity fields.
For the 33 RPSBs with both stellar and emission line kinematic measurements, 10 of them have misaligned gas and stellar kinematics, and a further 13 show other interaction features. Therefore, we conclude that $\gtrsim 50$\% of both CPSB and RPSB galaxies have evidence 
for kinematic or morphological disturbance. 
For the CPSB control samples, on average we find 1/31 galaxies with interaction evidence, and 3/31 galaxies with misaligned gas and stellar kinematics.  For the RPSB control samples, 
on average we find 1/37 misaligned galaxy and no interaction evidence. 
The obvious difference between PSB galaxies and their control samples suggests that mergers, interaction or recent gas accretion may plausibly cause the formation of PSB regions.



\section{Discussion}

The optical spectral features of post-starburst galaxies have been interpreted as a 
signature of an abrupt decrease in the star-formation activity, likely following 
a recent starburst. This means that PSB regions have rapidly quenched their 
star formation in the recent past and we can extrapolate to hypothesise that they are in transition from the blue cloud to 
the red sequence. Thus, they provide a unique insight into galaxy evolution and may 
offer a means of constraining the origin of the bimodal colour distribution of galaxies. 
However, the most direct and important problems regarding PSBs include: (1) what 
triggers the starburst at the beginning? (2) what mechanism quenches the starburst
on the short timescales (of order a single dynamical period) required to produce the PSB features?

To address these two questions, much work has been done in the past decade. 
Based on hydrodynamic simulations, a popular galaxy evolutionary picture has emerged in which two gas-rich disks merge, 
tidal torques channel gas to the galaxy centers and progress to heavily dust-obscured, central starbursts, 
which is coupled with SMBH fueling and subsequent expulsion of gas, leading to the development of quiescent spheroids \citep[e.g.][]{hopkins06}. The PSB phase can be fit into this picture as feedback clears out the leftover gas, both star formation and black hole growth are ceased, the galaxies pass through the post-starburst
phase before they become ``red and dead". 

Observational evidence supporting this picture includes the high level of morphological disturbance in local PSB galaxies \citep[Section \ref{sec:kin} and][]{zabludoff96, yang04, yang08, pawlik18}, and fast outflowing gas detected in the \mgii~2796, 2803 absorption lines of $z>0.5$ PSBs \citep[][Maltby et al. subm]{tremonti07}, which may be a fossil galactic wind launched at the QSO stage. However, the low stellar mass of local PSBs, as well as the absence of powerful AGN in the local Universe makes this scenario less plausible for our sample. Additionally, the presence of cold gas in many PSBs shows that the gas has not been effectively expelled \citep{rowlands15,french15}. The limited resolution of the simulations, and necessary implementation of sub-grid star formation and feedback recipes, means we are still not sure whether a violent major merger/interaction is absolutely necessary to produce a starburst or the strong Balmer absorption lines seen in post-starburst galaxies, or whether an AGN is necessary or able to halt the starburst rapidly enough to produce galaxy wide post-starburst features. 
Our catalogue of PSB regions in the galaxy-wide MaNGA survey allows us to address these questions from a new angle, compared to the single fibre surveys carried out in the past. With this new large and uniform IFU sample, we can get some new clues for understanding the origins of the PSB features.

For the CPSBs, the strong radial gradients in \d4n\ and \hda\ certainly support a scenario in which gas has flowed rapidly into the centre of the galaxy, leading to a strong central starburst. The high fraction ($\gtrsim 50$\%) of objects with misaligned gas-star kinematics or tidal features supports the idea that a merger has triggered the gas inflow in the majority of cases. The globally young mass-weighted age indicates a violent process has mixed the stars throughout the galaxy, and the significant decrease in stellar $v/\sigma$ compared to the control sample again implies a violent interaction or merger has occurred. It seems unlikely that less violent processes, such as misaligned gas accretion from a neighbouring dwarf or the cosmic web could contribute significantly to the population. The weak \Wha\ throughout the disk implies subsequent galaxy-wide quenching of the star formation, which could be caused by complete gas exhaustion, expulsion or some additional heating mechanism. While AGN feedback has been postulated as a plausible mechanism for global quenching, it is unclear whether it can cause such galaxy wide quenching in the relatively low mass PSB galaxies present in our local Universe sample. 

For the entirely new class of RPSBs, the strong radial gradient in \Wha\ compared to the control galaxies again suggests a recent strong inflow of gas to the central regions. 27\% of RPSBs show misaligned gas and star kinematics,
19\% have obvious bars, 30\% show interaction evidence like tidal tails, while the remainder are either face or edge-on making it difficult to identify bars or kinematic misalignment. All these mechanisms can lead to gas inflows \citep{lin17, chown19},
indicating that gas inflow is also key in the formation of the RPSBs. In contrast to the CPSBs, however, the RPSB galaxies have not (yet) suffered global quenching of the star formation. The outer regions are identified as PSBs due to their weak \Wha, however, typically the residual \Wha\ is still stronger than in the CPSBs. The presence of both RPSBs and IPSBs provides direct evidence that an AGN is not a necessary ingredient to rapidly quench a starburst and cause post-starburst features in galaxies. 

The question that naturally arises is whether the RPSBs are simply younger relatives of the CPSBs, probing different evolutionary phases of the same event. This would be consistent with the ongoing star formation in the central regions of RPSBs, and higher fraction with clear evidence for tidal tails and interaction which fades rapidly with time since the starburst \citep{pawlik18}. If this were true it would also be tempting to suggest a scenario in which either the starburst or the quenching progresses from the outer regions inwards, the latter option presumably ruling out AGN feedback as the quenching mechanism. However, two lines of evidence suggest against this hypothesis. Firstly, the two populations have very different star formation histories at all radii (as indicated by their mass-weighted ages), and secondly the CPSBs have lower stellar $v/\sigma$ at all radii. Neither of these observational features are alterable on timescales of a few 100\,Myr needed to shut down the central starburst in the RPSBs whilst retaining the strong Balmer absorption lines leading to the CPSB classification. We therefore conclude that the CPSB and RPSB galaxies are likely the product of different physical mechanisms.  

In the H$\alpha$ emission vs. H$\delta_{\rm A}$ absorption plane, the RPSBs can be separated into two types: type I appear to be undergoing a global shut down in star formation which could plausibly be caused by strangulation processes where the infall of fresh gas is shut off; type 2 appears to require a more complex process with the outer regions quenching first while the inner regions remain star forming, perhaps caused by ram pressure stripping or interactions/mergers. Why these mechanisms would coincide with the requisite gas inflows to cause the central starburst remains unclear. In order to pin down the quenching mechanisms in different types of PSB galaxies,  comparisons with detailed hydrodynamic simulations are clearly required.



\section{Summary}
We identify galaxies with PSB regions in the MaNGA survey, generating a sample of 31 central (CPSBs), 37 ring (RPSBs) and 292 irregular (IPSBs) post-starburst galaxies. This is the first time that we are able to search for PSB regions across the full galaxy area, rather than focus on the central PSB regions. With this large IFU post-starburst sample, there are several important results that can be summarized as:

\begin{enumerate}
\item Based on the global properties of the galaxies, we find that RPSBs are primarily located on the star forming main sequence while CPSBs are primarily located in the green valley. While a higher fraction of CPSBs have S$\acute{\rm e}$rsic index $n>3$ indicating pure spheroidal morphologies, CPSBs and RPSBs cover a wide range in $n$, showing that both types are hosted by galaxies with diverse morphologies. 

\item Both CPSBs and RPSBs have positive gradients in \d4n, indicating younger stellar populations in the central regions. This is different to control samples, which have flat or negative gradients. 


\item While the CPSBs have suppressed star formation throughout their bulge and disk, and clear evidence of rapid quenching in the central regions, the RPSBs only show clear evidence of recently rapidly suppressed star formation in their outer regions and ongoing central star formation/starburst. 

\item The different radial profiles in mass-weighted age and stellar $v/\sigma$ indicate that CPSBs and RPSBs are not simply different evolutionary stages of the same event, rather that CPSB galaxies are caused by a significant disruptive event, while RPSB galaxies are more likely caused by disruption of gas fuelling to the outer regions. 

\item Compared to the control samples, both CPSB and RPSB galaxies show a higher fraction of interactions/mergers, misaligned gas or bars that might be the cause of the gas inflows. 

\item The presence of both RPSBs and IPSBs provide direct evidence that an AGN is not a necessary ingredient to cease starburst. 

\item The wide range in observed radial profiles of H$\alpha$ emission vs. \hda\ absorption in the RPSBs indicate that multiple processes may be responsible for their shut off in star formation, such as strangulation processes where the infall of fresh gas is shut off leading to a global shut down in star formation, or ram pressure stripping or interactions/mergers leading to the rapid quenching of the outer regions while the inner regions remain star forming.
\end{enumerate}

\section*{acknowledgements}
Y. C acknowledges support from the National Key R\&D Program of China (No. 2017YFA0402700), the National Natural Science Foundation of China (NSFC grants 11573013, 11733002, 11922302).
VW and KR acknowledge support of the European Research Council via the award of a starting grant (SEDMorph; P.I. V. Wild). CT acknowledges NSF CAREER Award AST-1554877. 
DB is partly supported by RSCF grant 19-12-00145. RR thanks CNPq, FAPERGS and CAPES for partially funding this project. 
We thank MSc student John Proctor for carefully proof reading an early manuscript. 

Funding for the Sloan Digital Sky Survey IV has been provided by the Alfred P.
Sloan Foundation, the U.S. Department of Energy Office of Science, and the Participating Institutions.
SDSS-IV acknowledges support and resources from the Center for High-Performance Computing at
the University of Utah. The SDSS web site is www.sdss.org.

SDSS-IV is managed by the Astrophysical Research Consortium for the Participating Institutions of the SDSS Collaboration 
including the Brazilian Participation Group, the Carnegie Institution for Science, Carnegie Mellon University, the Chilean Participation 
Group, the French Participation Group, Harvard-Smithsonian Center for Astrophysics, Instituto de Astrof\'isica de Canarias, The Johns 
Hopkins University, Kavli Institute for the Physics and Mathematics of the Universe (IPMU) / University of Tokyo, Lawrence Berkeley 
National Laboratory, Leibniz Institut f\"ur Astrophysik Potsdam (AIP), Max-Planck-Institut f\"ur Astronomie (MPIA Heidelberg), 
Max-Planck-Institut f\"ur Astrophysik (MPA Garching), Max-Planck-Institut f\"ur Extraterrestrische Physik (MPE), National 
Astronomical Observatories of China, New Mexico State University, New York University, University of Notre Dame, 
Observat\'ario Nacional / MCTI, The Ohio State University, Pennsylvania State University, Shanghai Astronomical Observatory, 
United Kingdom Participation Group, Universidad Nacional Aut\'onoma de M\'exico, University of Arizona, University of Colorado Boulder, 
University of Oxford, University of Portsmouth, University of Utah, University of Virginia, University of Washington, University of Wisconsin, 
Vanderbilt University, and Yale University.

\bibliographystyle{mn2e}

\bibliography{psb}

\end{document}